\documentclass[fleqn,twoside]{article}
\usepackage{espcrc2}
\usepackage{times}
\usepackage{psfig}

\newcommand{\AmS}{{\protect\the\textfont2
  A\kern-.1667em\lower.5ex\hbox{M}\kern-.125emS}}

\newcommand\cxo{{\em Chandra}}
\newcommand\chandra{{\em Chandra}}
\newcommand\asca{{\em ASCA}}
\newcommand\xmm{{\em XMM-Newton}}
\newcommand\sax{{\em BeppoSAX}}
\newcommand\rxte{{\em RXTE}}
\newcommand\comptel{{\em CGRO-Comptel}}
\newcommand\osse{{\em CGRO-OSSE}}

\newcommand\casa{Cas~A}
\newcommand\rxjSNR{{RXJ~1713.7-3946}}

\newcommand\msun{{M$_{\odot}$}}
\newcommand\net{$n_{\rm e}t$}
\newcommand\tiff{{$^{44}$Ti}}
\newcommand\scff{{$^{44}$Sc}}
\newcommand\caff{{$^{44}$Ca}}
\newcommand\nifs{{$^{56}$Ni}}

\newcommand{\fluxunit}{{ph\,cm$^{-2}$s$^{-1}$}}

\newcommand\arcmin{\mbox{$^\prime$}}%

\def\nat{{Nat\,}}
\def\adspr{{AdSpR}}
\def\apj{{ApJ\,}}
\def\apjl{{ApJL\,}}
\def\apjs{{ApJS\,}}
\def\aap{{A\&A\,}}
\def\aaps{{A\&AS\,}}
\def\pasp{{PASP\,}}
\def\pasj{{PASJ\,}}

\title{A Review of X-ray Observations of Supernova Remnants}

\author{Jacco Vink
\address{Columbia Astrophysics Laboratory, Columbia University, 
New York, NY, USA}
\address{Present address: SRON National Institute for Space Research,
Sorbonnelaan 2, 3584 CA Utrecht, Netherlands}
\thanks{Chandra fellow}}

\begin{document}

\begin{abstract}
I present a review of X-ray observations of supernova remnants with
an emphasis on shell-type remnants.
The topics discussed are the observation of fresh nucleosynthesis products,
shock heating and cosmic ray acceleration.
\end{abstract}

\maketitle

\section{INTRODUCTION}

Although radio and optical observations have played an important role
in identifying and characterizing supernova remnants (SNRs), 
most of the shock heated material is primarily visible in X-rays,
having temperatures of typically $10^6-10^7$~K. 
The emission mechanisms are bremsstrahlung and line radiation,
but a number of remnants are now known to emit X-ray synchrotron radiation
from ultra-relativistic electrons, as well. This was already known for
pulsar wind nebula dominated, or Crab-like remnants, which 
fall outside the scope of this review, but also shell-type remnants
like SN 1006 emit synchrotron radiation from electrons
accelerated by the blast wave \cite{koyama95}.

The X-ray line emission of SNRs is in particular interesting, as it
allows the study of the most abundant metals, the so called
$\alpha$-elements, which prominently feature in the X-ray spectra
of young SNRs like \casa, Tycho (SN 1576) and Kepler (SN 1064).
This line emission, and mass estimates based on the X-ray emission,
give important information on the
nucleosynthesis yields of the explosions, which can then
be compared to theoretical supernova (SN) models \cite{ww95,nomoto97}.

The line and bremsstrahlung emission give us a window on the past of
the SNR: its progenitor and the SN explosion.
X-ray synchrotron radiation, on the other hand,
is interesting as it provides a window on cosmic ray shock acceleration.
SNRs are thought to be the primary source of cosmic rays up to, and possibly
beyond, energies of $10^{15.5}$~eV, at which energy the spectrum
has a break, usually referred to as ``the knee'',
However,
direct evidence for a SNR origin for cosmic ray particles other than electrons
is lacking.
For a long time the observational support for a SNR origin of cosmic rays
was the detection of radio synchrotron emission from SNRs, revealing
the presence of cosmic ray electrons up to energies of a few GeV.
Recently the detection of X-ray synchrotron radiation from SNRs have
substantially extended this to  $\sim100$~TeV \cite{koyama95,reynolds99}.

In this review I will discuss recent findings concerning both the study
of fresh nucleosynthesis and cosmic ray acceleration in SNRs,
emphasizing the role of X-ray observations, including
results based on \sax\ observations.

\begin{figure}
\psfig{figure=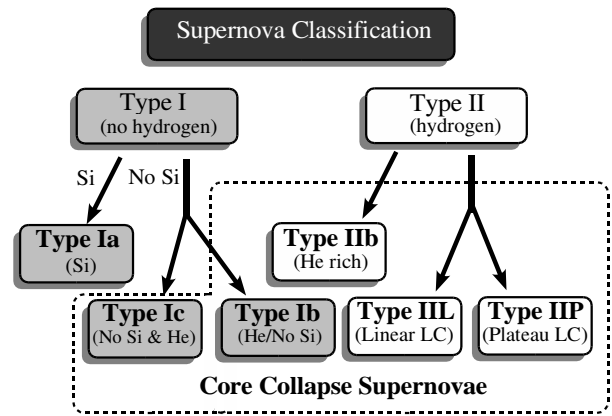,width=0.5\textwidth}
\caption{The supernova classification scheme, which is
based on optical spectroscopy and light curve (LC) shape.\label{scheme}}
\end{figure}

\section{SUPERNOVA REMNANTS AND THEIR PROGENITORS}
Two recent astrophysical developments have boosted the interest
in supernovae (SNe): the use of Type Ia SNe as standard candles to
study the geometry of the Universe (e.g. \cite{tonry03}), and the discovery
that gamma-ray bursts (GRBs) may be associated with Type Ib/c 
SNe (e.g. \cite{stanek03}).

Type Ia and Type Ib/c SNe are very different objects 
(see Fig.~\ref{scheme}): Type Ia SNe are probably caused by the
thermonuclear disruption of a C-O white dwarf following mass accretion, 
but the nature of the binary system, nor the disruption mechanism
(detonation, deflagration, delayed detonation) are yet known
\cite{branch95}.
Type Ib/c and Type II SNe, on the other hand, are core
collapse SNe of massive
stars, i.e. $>8$~\msun\ while on the main sequence (MS).
The classification is for historical reasons confusing.
Only since two decades are Type Ib/c recognized as
a special subclass, like Type Ia their optical spectra show no
H absorption, but for Type Ibc the likely reason is 
the loss of the hydrogen envelope of a massive star,
either due to binary mass transfer or
due to a strong stellar wind.

The energy source for the explosion of core collapse SNe is the
gravitational collapse of the stellar core. The stellar core itself
will leave a neutron star or black hole behind, and the energy for the 
explosion comes from the release of gravitational binding energy. 
The energy is primarily
released in the form of neutrinos, but to order 1\% 
is kinetic energy. The SN-GRB connection has led to a new interest
in the explosion mechanism itself. A large fraction of the GRB energy
must be in the form of jets, and a natural question then is how these
jets are formed and whether jet-formation plays a role in the
explosion of core-collapse SNe in general.

\begin{figure*}
\centerline{
\psfig{figure=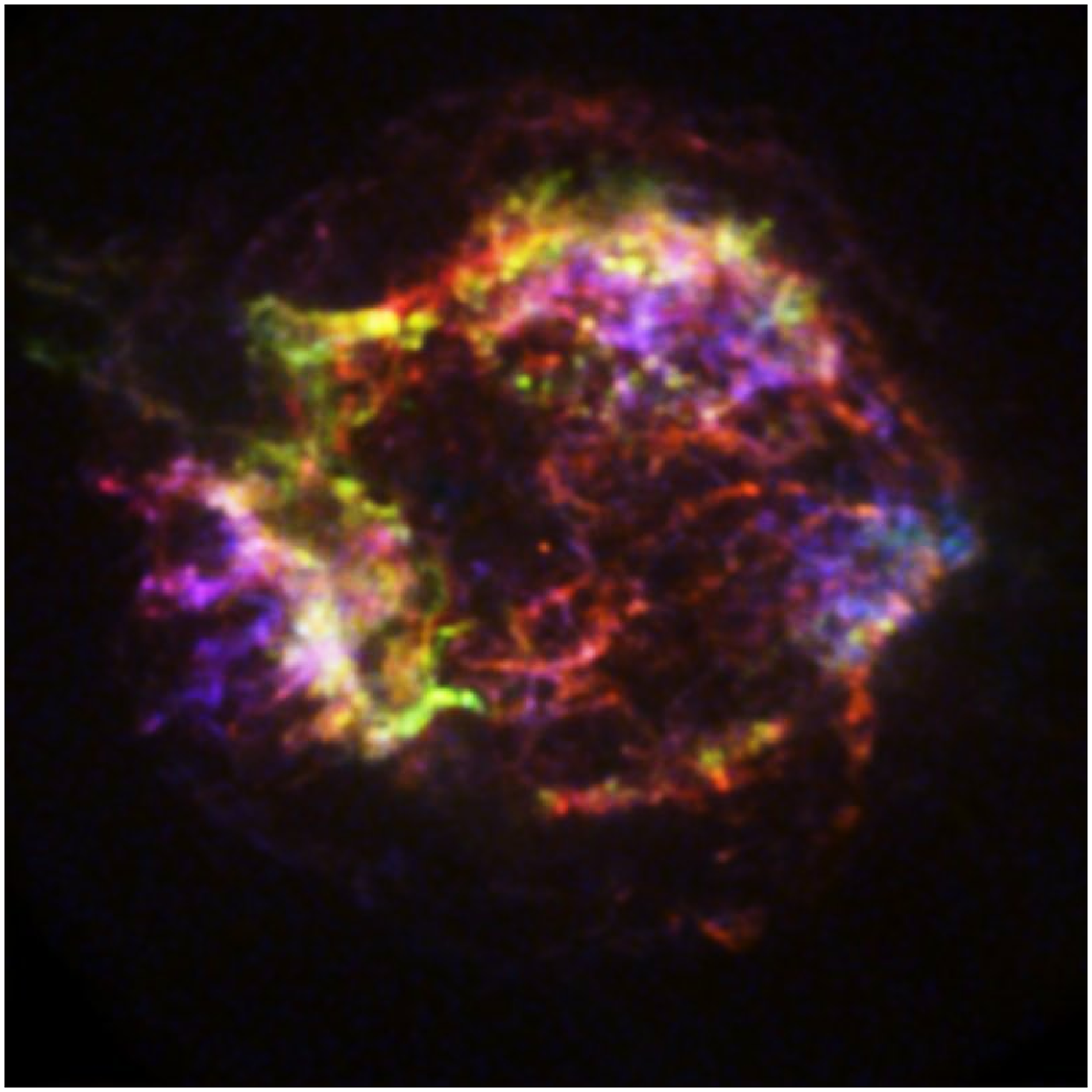,width=0.36\textwidth}
\psfig{figure=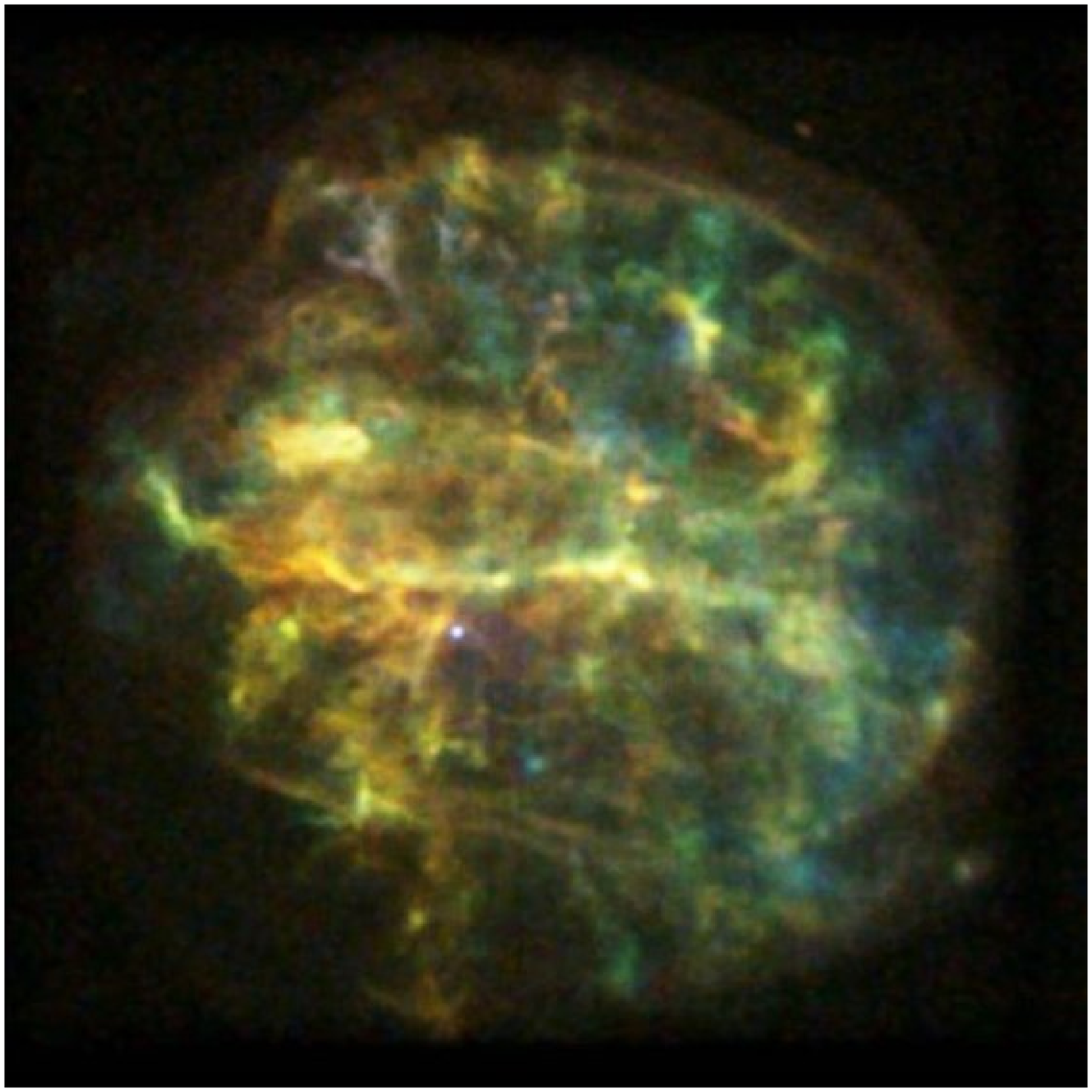,width=0.36\textwidth}
}
\centerline{
\psfig{figure=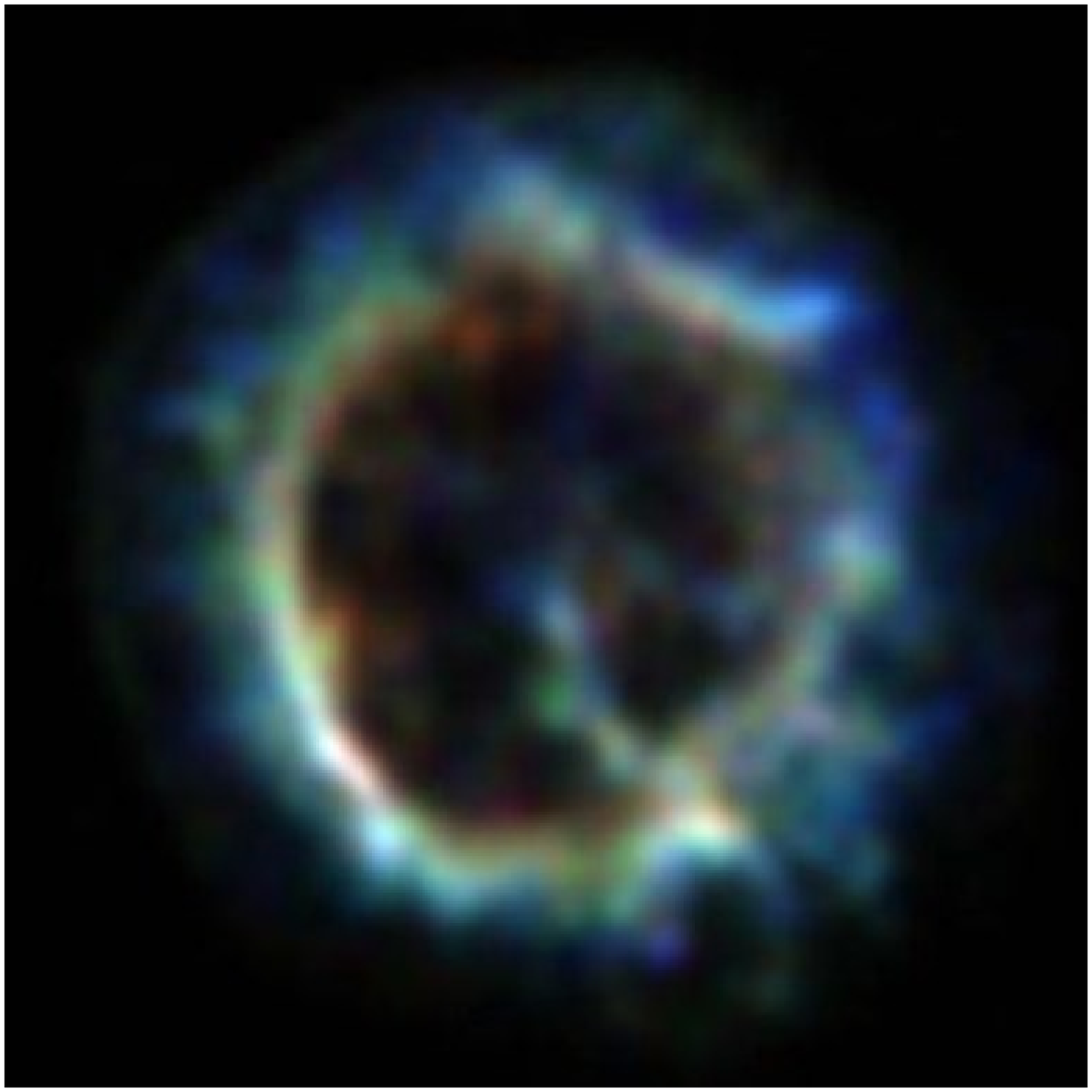,width=0.36\textwidth}
\psfig{figure=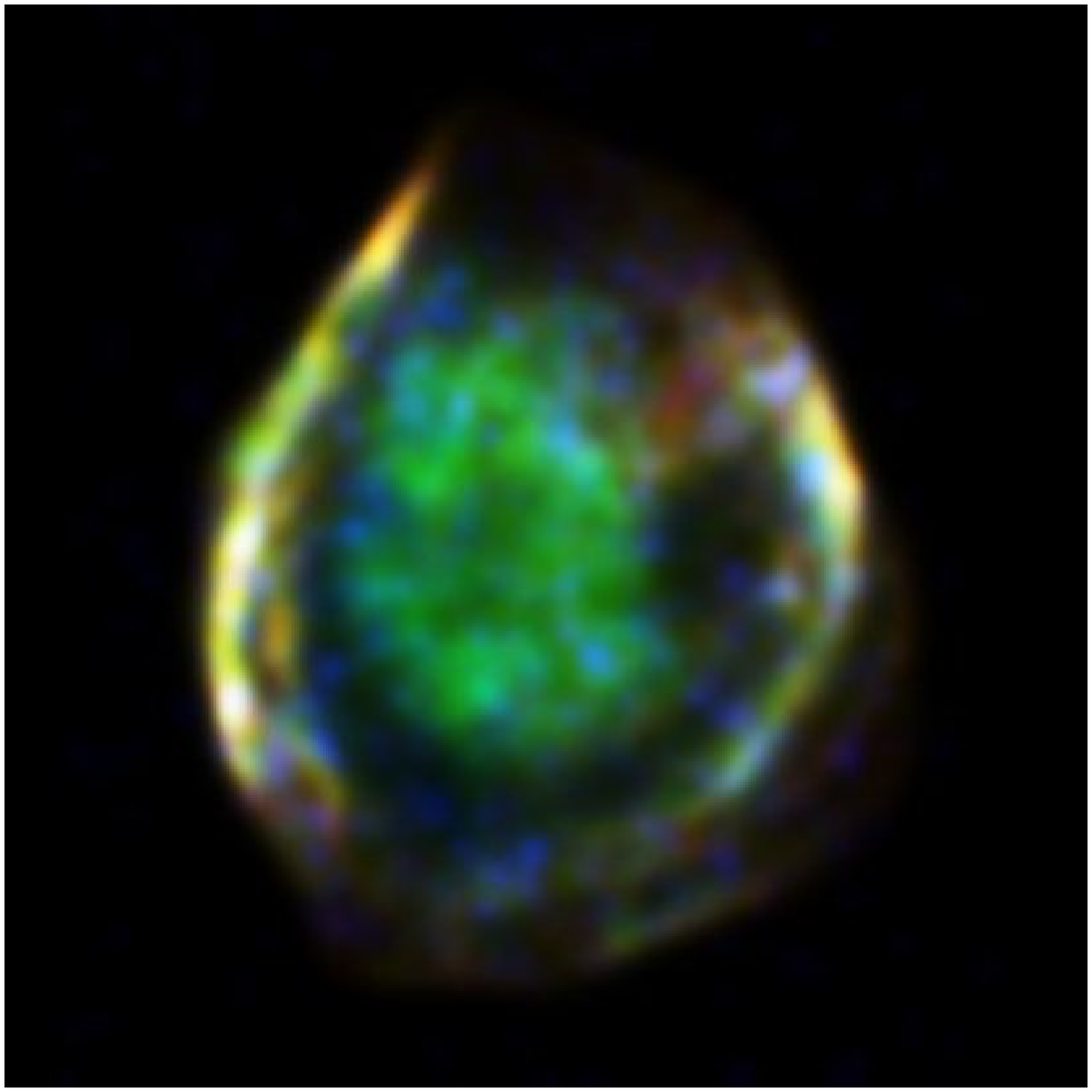,width=0.36\textwidth}
}
\centerline{
\psfig{figure=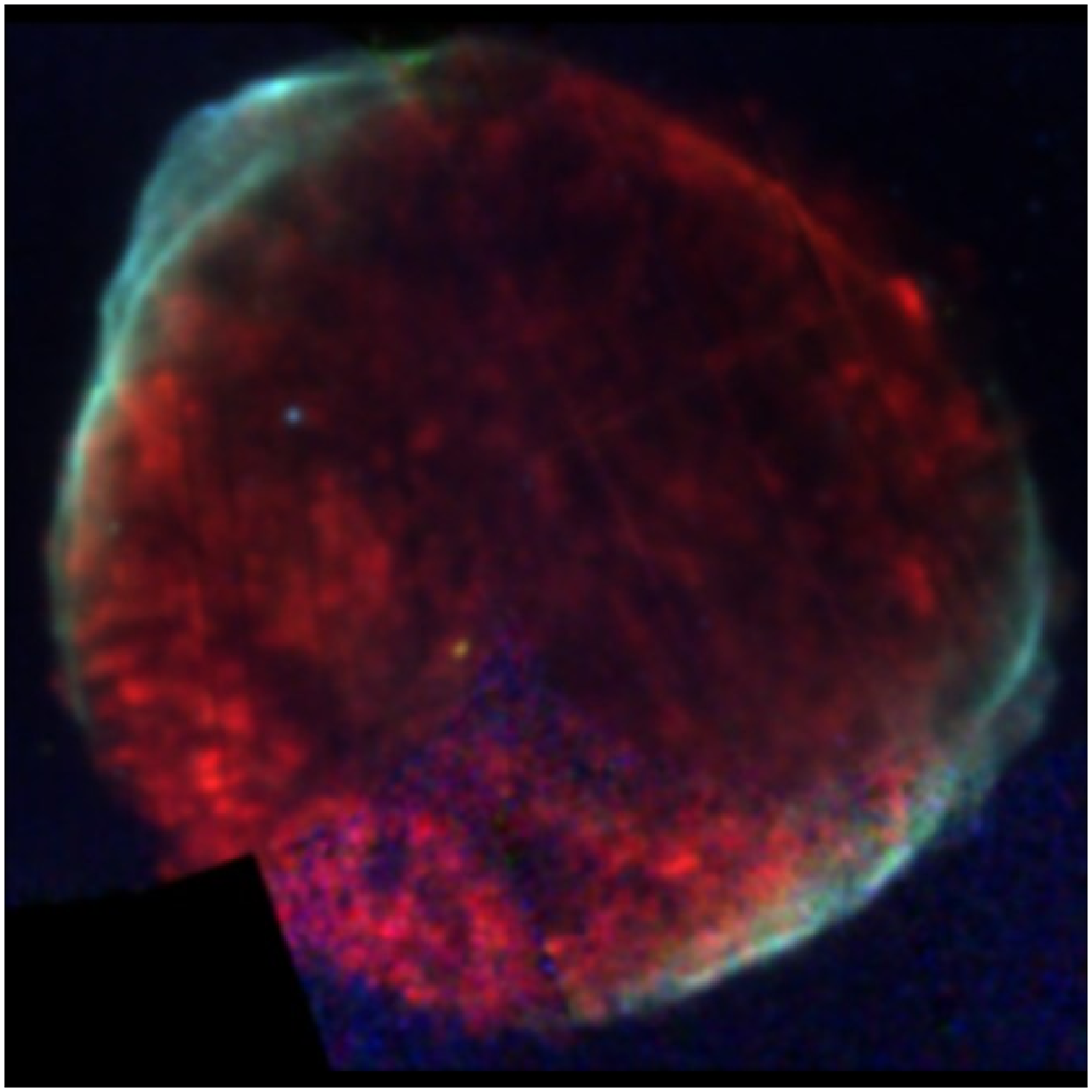,width=0.36\textwidth}
\psfig{figure=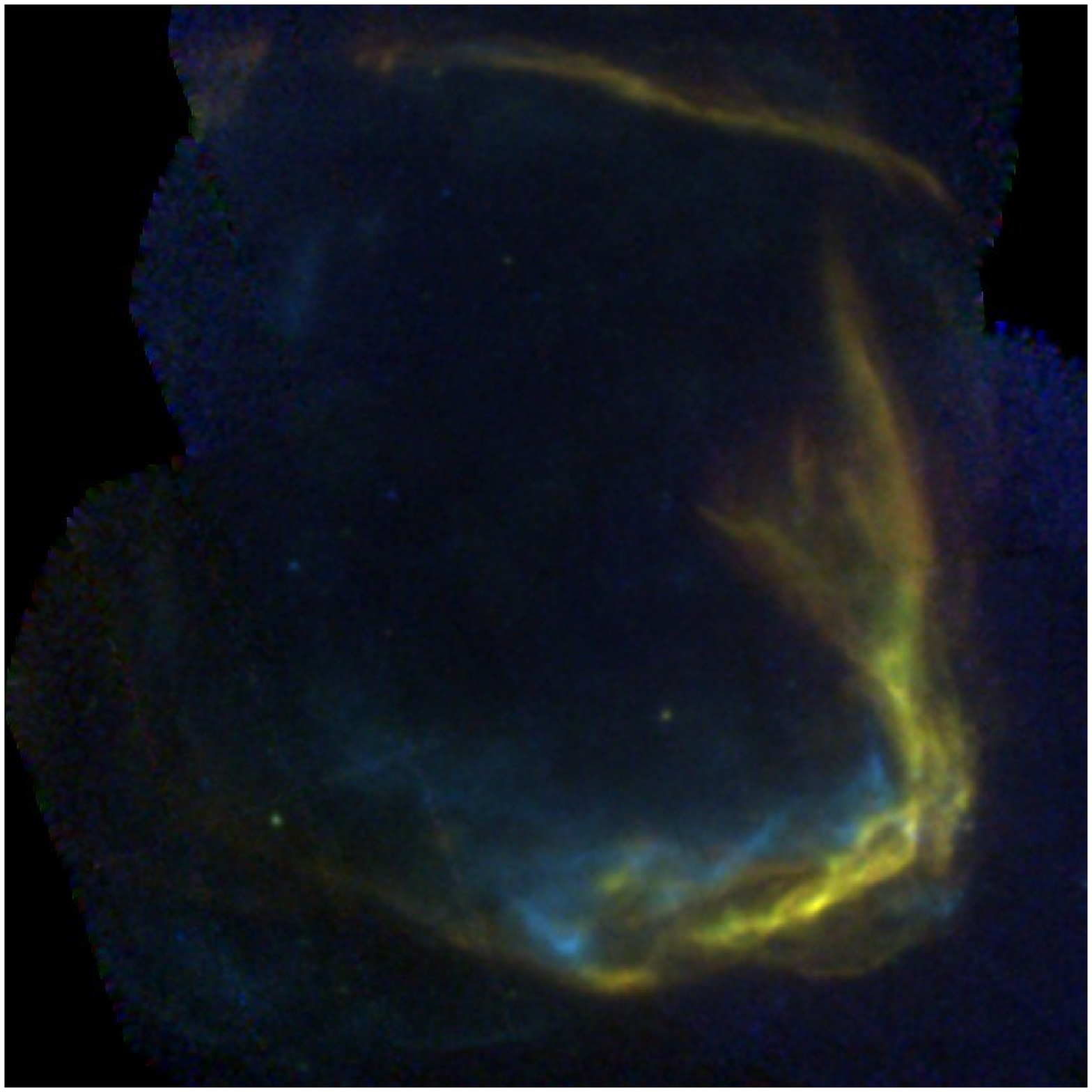,width=0.36\textwidth}
}
\caption{
\chandra\ and \xmm\ (bottom images) images of supernova remnants.
From left to right, top to bottom: 
Cas A (Mg XII, Si XIII), 
G292.2-1.8 (OVII/VIII, Ne, Si XIII),
E0102-7219 (OVII, OVIII, Ne/Mg),
Dem L71 (O,Fe L, Si), SN 1006 (O VII, 0.7-2 keV, 2-7 keV),
and RCW 86 (O VII, 0.7-2 keV, 2-7 keV).
The labels in parenthesis refer to the energy bands used
for the RGB colors.
\label{color}}
\end{figure*}

\begin{figure*}
\centerline{
\psfig{figure=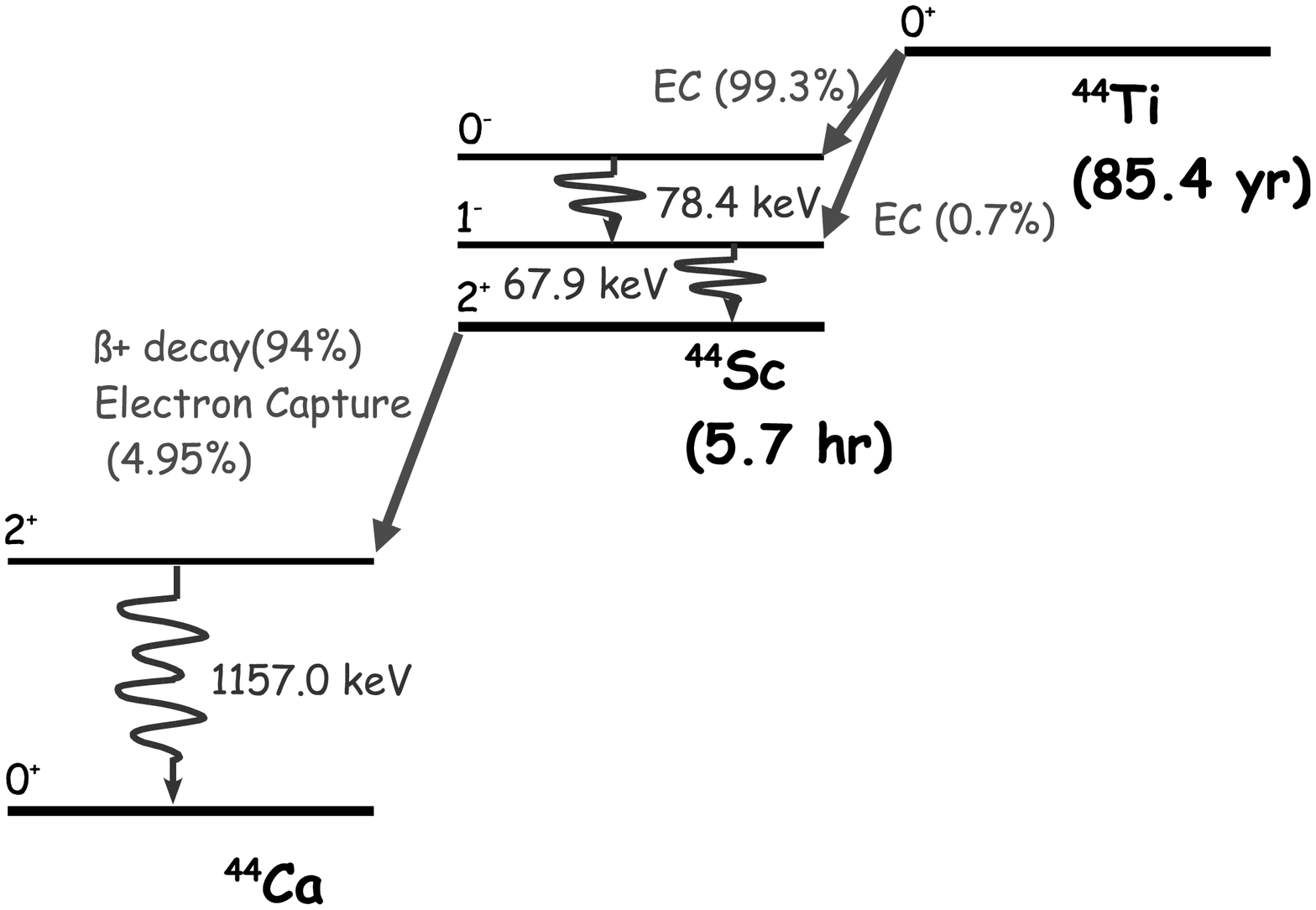,width=0.5\textwidth}
\psfig{figure=jvink1_f3b.eps,angle=-90,width=0.5\textwidth}}
\caption{
The decay scheme of \tiff\ (left) and the \sax-PDS spectrum of
\casa, showing the presence of
the 68~keV and 78~keV (\scff) decay lines (right).
The various contributions to the spectrum are indicated separately,
the dotted line indicating the \scff\ contribution. 
The low level components between 20-60~keV is instrumental (escape
peaks of the line emission).
\label{ti44}}
\end{figure*}

\subsection{Core collapse remnants}
Both core-collapse and Type Ia SNe release $\sim10^{51}$~erg
in kinetic energy. It is therefore sometimes difficult to distinguish
remnants from both types of explosions.
However, the presence of a pulsar clearly identifies a remnant
as a core collapse remnant. In other cases we have to rely on abundance
patterns, if relatively unmixed ejecta are present, 
and circumstantial evidence such as the proximity of
an OB association.
Clearly identifiable ejecta are found in young remnants, but
have also been found
in SNRs of several thousand years old; e.g. the fast moving
ejecta fragments associated with the Vela remnant \cite{aschenbach95}, or
Si-rich, X-ray emitting material from the center of the
Cygnus Loop \cite{miyata98}.

Younger remnants are useful to obtain a more complete understanding
of nucleosynthesis yields. Especially the so called O-rich SNRs
are important, as they are likely to be remnants of the most massive stars,
which have a higher O yield. 
These remnants are therefore possibly remnants of Type Ib/c SNe.
Among them are the young SNRs \casa,  G292.0+1.8 (MSH11-54) and E0102-7219 
(Fig.~\ref{color}).
Initially their identification as O-rich was based on optical line emission,
but X-ray spectroscopy confirms that they are O-rich.
Interestingly, although they are probably remnants of very massive
stars, which are thought to produce black holes rather than neutron stars
\cite{heger03}, G292.0+1.8 harbors a 135~ms pulsar \cite{camillo02}
and \casa\ a neutron star candidate in the form of a
non-pulsating point source, similar to another O-rich
remnant, Puppis A \cite{pavlov02}.
\casa\ may have been a 18-25~\msun\ MS star, consistent with
the presence of a neutron star, but a recent analysis of \chandra\ data 
of G292.0+1.8
led \cite{gonzalez03} to the conclusion that it is the remnant of a 
30-40~\msun\ MS star,
which according to \cite{heger03} should have produced a black hole.

\casa\ is the youngest galactic remnant, and one of the best studied ones
(\cite{vink04}, for a review).  
It is also the remnant that was observed very extensively with \sax\ 
\cite{favata97,vink99,vink00b,maccarone01,vink01a}.
One remarkable feature observed with \sax\ was the fact that the Fe-K line
emission peaked in the southeast,
outside the main Si-rich ejecta shell \cite{vink99}.
\chandra\ has revealed
that the Fe-K emission is not coming from shock heated circumstellar material,
but from Fe-rich ejecta knots \cite{hughes00a}.
This is quite remarkable, because, according to
standard SN models, the Fe-rich shell
lies inside the Si and O-rich shells.
Some of the knots seem to be almost pure Fe, and have been identified
as $\alpha$-rich freeze-out products \cite{hwang03}. This
process arises if during the explosion very hot material, $5\times10^9$~K,
rapidly expands.
In the northern part of \casa\ other Fe-rich material is found inside the main
shell, but an X-ray Doppler study, based on \xmm\ observations,
reveals that the Fe in this region 
is moving faster than Si and S. The position of Fe inside the northern shell
is therefore likely to be a projection effect \cite{willingale02}.

The \xmm\ study also confirms the large deviations from spherical symmetry
in the expansion of Si, S and Fe, with the southeastern part having a bulk 
redshift and the northern part a bulk blueshift.
The asymmetries are likely due to explosion asymmetries in the kinematics of
the ejecta from the core as the forward
shock, which is more indicative of the circumstellar medium and 
is marked by a narrow X-ray synchrotron emitting filament 
\cite{gotthelf01a}, is
remarkably circular. An aspherical ejecta expansion
has also been found for the SMC O-rich SNR
E0102-7219 with \chandra-HETG high resolution spectroscopy \cite{flanagan01}.

Keeping in mind the GRB jets, one may wonder
wonder whether the asymmetries in these likely Type Ib systems are
related to the asymmetries in the much more energetic GRBs.
\casa\ has an interesting jet-like structure in the northeast, with a
possible weaker counterpart in the west \cite{vink04}. However,
it consists of Si-rich material, and is therefore unlikely to have formed
close to the collapsing core. The Fe-rich core material on the other
hand is clearly aspherical, but its kinematics do not reveal
any bipolarity. This is not necessarily in disagreement with a scenario
in which a SN explodes with a collapsar-like mechanism, i.e. an accretion
powered SN, 
because collapsar models show that \nifs, which decays to Fe, 
is not produced in the jets but in a ``wind''
in between the jets and the accretion disk \cite{macfadyen03}.
On the other hand the latest numerical simulations of a conventional
core collapse also indicates the formation of high velocity
Ni-rich material \cite{kifonidis03}.

\nifs\ nucleosynthesis in an environment in which the density
drops rapidly due to expansion gives rise to $\alpha$-rich freeze out 
\cite{arnett96}. The synthesis of radio-active \tiff\ is a very
characteristic byproduct of this process. 
\tiff\ decays in $\sim85$~yr into \scff, which decays
to \caff\ (Fig.~\ref{ti44}). 
Interestingly, \casa\ is one of only two objects in which
\tiff\ has been detected \cite{iyudin94,iyudin98}. 
The characteristic \caff\ at 1157~keV was detected
by \comptel\ \cite{iyudin94} with a flux of $7\times10^{-5}$~\fluxunit.
For some time, however, this detection was not consistent with
the non-detection of the \scff\ lines at 68~keV and 78~keV by
\osse\ and the high energy experiments on board 
\rxte, and \sax\ \cite{the96,rothschild98,vink00b}.
Finally, however, a 500~ks \sax\ observation of \casa\ resulted in a better
than 3$\sigma$\ detection of those lines (Fig.~\ref{ti44}) 
\cite{vink01a,vink03a} with a flux of $(3.3\pm0.3)\times10^{-5}$~\fluxunit,
with some uncertainty due to the uncertain shape of the hard X-ray continuum.
Assuming an age for \casa\ of 320~yr and a distance of 3.4~kpc \cite{reed95},
the combined \comptel\ and \sax\ detections translate
into an initial \tiff\ mass of $(1.8\pm0.3)\times10^{-4}$~\msun.
This is higher than expected based on the standard nucleosynthesis models
\cite{ww95}, but consistent with either large scale explosion asymmetries or
a higher explosion energy ($2\times10^{51}$~erg instead of $10^{51}$~erg,
see also \cite{laming03}).
The detection of \tiff, together with reasonably reliable
yields of other elements \cite{vink96,willingale03}, make \casa\ one of
the best examples of how X-rays helps us to form a better picture
of core collapse SNe.

\begin{figure}
    \psfig{figure=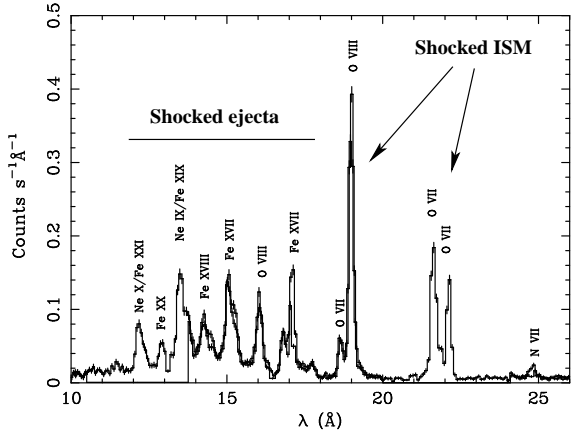,width=\columnwidth}
\caption{
The high resolution \xmm\ RGS spectrum of Dem L71 \cite{vanderheyden03}.
\label{typeIa}}
\end{figure}

\subsection{Type Ia remnants}
In order for a C-O white dwarf to be disrupted and eject material with
a high velocity more than 0.5~\msun\ of C-O has to be burned into
\nifs.
One can therefore expect that young Type Ia SNRs copiously emit
Fe line emission, either Fe-L emission
in the energy range from 0.8-1.2~keV (Fig.~\ref{typeIa}), or 
Fe-K emission from 6.4-7 keV.

The Fe-L emission consists of many lines giving rise to a bump in 
X-ray spectra observed with CCD detectors such as found on \asca, 
\xmm, and \cxo.
In \cite{hughes95} it was observed that LMC remnants fall into two classes,
those with prominent Fe-L emission and those without. 
The Fe-L dominated remnants are likely Type Ia remnants.
However, one should be careful in applying this method. For example,
Kepler's remnant would be accepted as a Type Ia remnant, whereas
optical spectroscopy of Kepler shows the presence of N overabundance
suggesting a core collapse remnant \cite{bandiera91}.
Another example is SNR Tycho (SN 1572), 
which is generally accepted to be a Type Ia remnant,
but it lacks prominent Fe-L emission.
However, Tycho's X-ray spectrum  does show Fe-K emission, which
peaks at a smaller radius than the spectral lines of other elements.
This clearly indicates that there is an elemental stratification,
just as expected for Type Ia remnants, with an inner ejecta layer
consisting of Fe and the outer ejecta of mid-Z elements 
\cite{hwang97,hwang02}.
The lack of Fe-L emission from
Tycho can be attributed to the low ionization age
of the plasma.\footnote{
SNR plasmas are often out of ionization equilibrium, due to
a combination of a low electron density, $n_{\rm e}$, and 
a short time, $t$, since the plasma was heated.}

A similar lack of Fe-L emission from another historical Type Ia remnant, 
SN 1006 (Fig.~\ref{color}),
can also be attributed to the low ionization age
of \net $\sim 2\times10^{9}$~cm$^{-3}$s \cite{vink99,dyer01,long03}.
An overabundance of Fe was reported in \cite{dyer01} and \cite{long03} 
but high resolution spectroscopy with
the \xmm\ RGS instruments did not reveal any signatures of Fe-L emission
\cite{vink03b}. Abundance determinations for SN 1006
of O, Ne, Mg, and Si are more reliable, as
those lines are clearly visible in CCD spectra.
So any low spectral resolution
X-ray Fe abundance measurements of SN 1006 should be regarded with caution,
especially since Fe is in an ionization stage ($<$ Fe XVII),
which produces almost no Fe-L emission.
Another reason to distrust Fe abundance measurements in SN~1006 is that optical
absorption spectra toward a bright UV background star reveals that
the reverse shock has probably not yet reached the Fe-rich ejecta
\cite{hamilton97}. 
SN 1006 is in that sense both dynamically and spectroscopically
younger than Tycho's remnant, a result of its evolution
in a low density environment.

In order to see all Fe one has to observe
dynamically older remnants, in which the reverse shock has heated 
almost all ejecta. A good example is Dem L71 \cite{hughes03}.
\cxo\ images in the energy bands of O and Fe-L show that most of the O VII/VIII
emission comes from a narrow shell, but most Fe-L line emission comes
from the center of the remnant 
(green in Fig.~\ref{color}, see also Fig.~\ref{typeIa}).
In fact, assuming that the plasma inside the shell consists of pure Fe,
the estimated Fe mass is $\sim$0.8~\msun \cite{hughes03,vanderheyden03}, 
consistent with Type Ia nucleosynthesis models \cite{nomoto97}.
In general, assuming that a plasma consists of a pure metal plasma, with
no or little H, will give rise to a lower overall mass estimate and
a higher mass estimate for that particular metal. Critical to this calculation
is the source of the exciting electrons.
Although, most of the Fe ejecta seems to be shocked, Dem L71 is less suitable
for obtaining the overall ejecta abundances, as the outer C-O, and Si
layers are likely to be mixed with the ISM.
Therefore, advances in our understanding of Type Ia SNe should probably
come from combining studies of young and old Type Ia remnants.

\section{SUPERNOVA REMNANTS AND COLLISIONLESS SHOCK PHYSICS}

Supernovae drive shock waves through the ISM that
remain visible for 10,000's of years in the form of SNRs. 
As the ISM is very tenuous,
the particle mean free path is larger than the typical shock width.
That a shock forms at all is due to the fact that heating occurs
not through two body interaction, but due to the coupling of particles with
self-generated plasma waves.
These so-called collisionless shocks 
are interesting from a physical point of view,
since the detailed shock heating process is unknown, and the
process can only be studied in the rarefied media 
in space.

\subsection{Shock heating}
\begin{figure}
\psfig{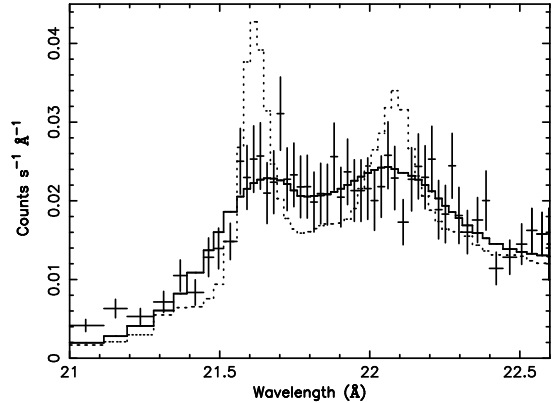}
\caption{
\xmm\ RGS high resolution spectrum of the bright compact knot at
the northwestern edge of the remnant. The dotted line shows the best
emission model excluding, the solid
line including thermal Doppler broadening \cite{vink03b}.
\label{sn1006}}
\end{figure}

One question considering collisionless shocks is whether they
heat each particle species according to its own Rankine-Hugoniot
equation\footnote{$kT_i = 3/16 m_i v_s^2$, 
for high Mach number shocks and species $i$, without additional energy sinks,
such as CR acceleration.},
or whether different particles are quickly
equilibrated by the collisionless heating process 
(i.e. $kT_{\rm e} = kT_{\rm H} = kT_{\rm He}$, etc.).
The first evidence for non-equilibration of temperatures was 
that UV line emission from SN 1006 indicates
that H, C, and O have approximately the same line width, instead of 
a line width inversely proportional to the particle mass, as can be expected
for an equilibrated shock \cite{laming96}.
Other evidence came from modeling the width of H$\alpha$\ line emission
in comparison with the narrow to broad line H$\alpha$ 
ratio observed in a number
of SNRs, which seem to indicate that low Mach number shocks are fully
equilibrated, whereas high Mach number shocks are not \cite{rakowski03}.
A recent measurement of the O VII line widths of a compact knot in SN 1006
indicate a high O VII temperature of
$\sim 350-700$~keV compared to an electron temperature of 
$\sim 1.5$~keV (Fig.~\ref{sn1006}, \ref{color}) \cite{vink03b}.

A remaining question is whether the shock heating process does always produce
a Maxwellian particle distribution and what the influence is of cosmic ray
acceleration on the temperature(s),
as cosmic rays provide an additional heat sink, or even lead to heat loss
as a result of CR escape. This process may have been observed in
E0102-7219, which showed an electron temperature too cold compared
to the observed expansion velocity, even if the non-equilibration of
electron and proton temperature was taken into account \cite{hughes00b}.

\begin{figure*}
\centerline{
  \psfig{figure=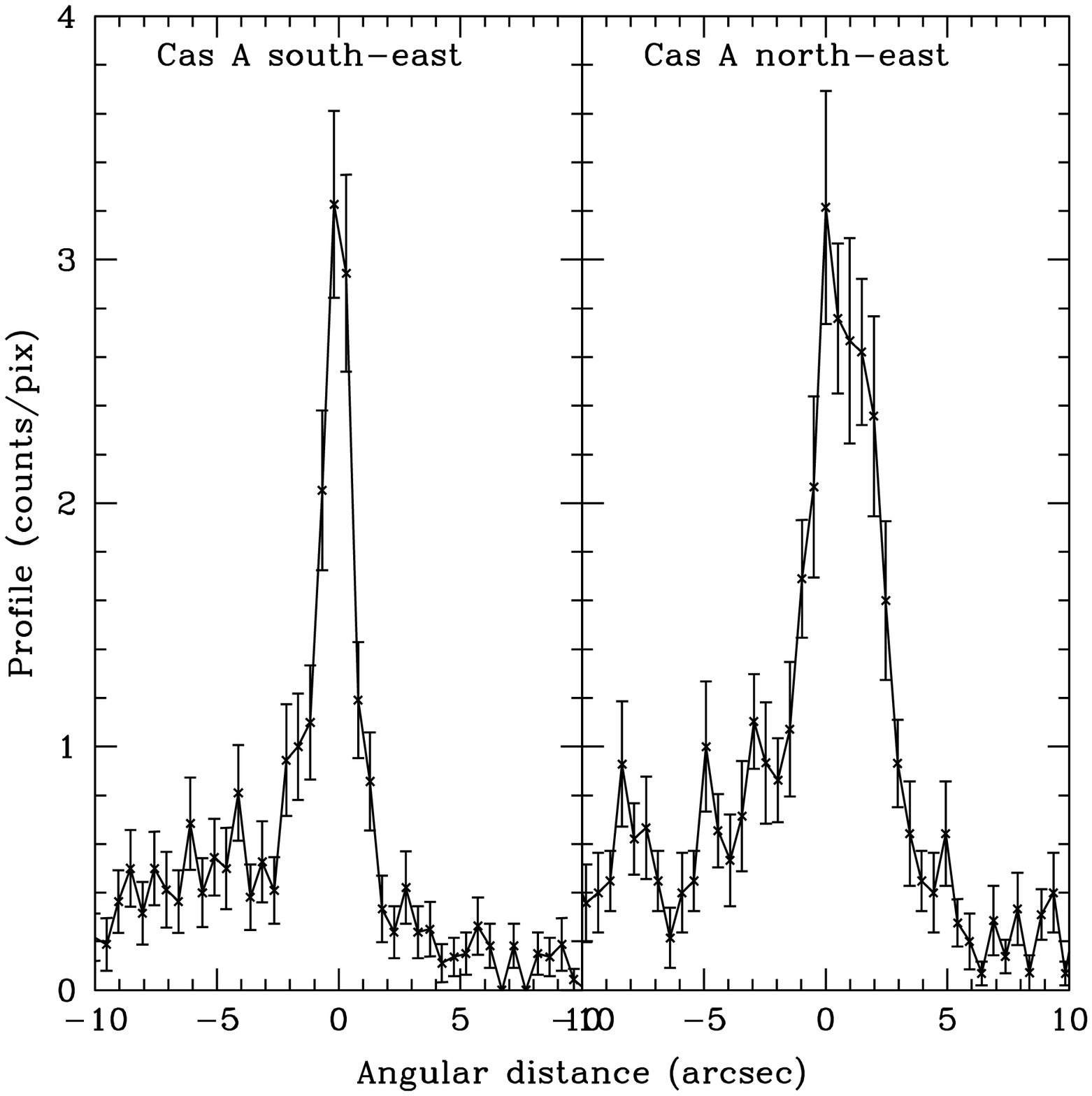,width=0.45\textwidth}
  \psfig{figure=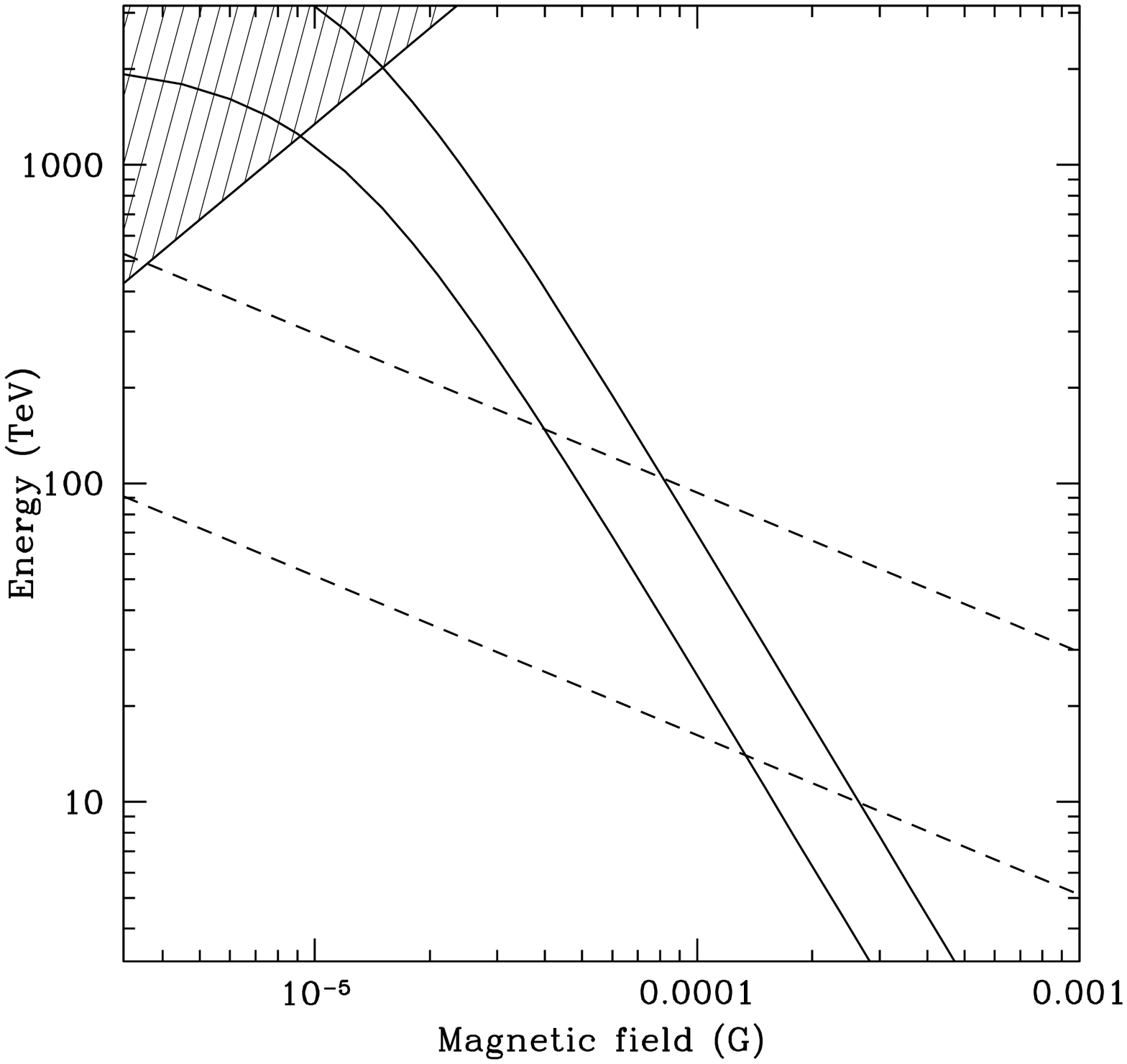,width=0.45\textwidth}
}
\caption{Left: Profiles across \casa's continuum rims (\chandra\ data).
The widths can be converted to a combination of magnetic field ($B$)
and electron energy ($E$), assuming that the width is determined by
synchrotron losses (curved lines, right).
Together with the observed photon energy (straight lines)
this gives approximate  $E$ and $B$ estimates
(after \cite{vink03a}).
\label{casarims}}
\end{figure*}

\subsection{Particle acceleration}

SNRs are considered to be the dominant source of cosmic rays up to
the ``knee''. The reason is that the likely galactic SN rate of 1-2 per century
is sufficient to maintain the observed cosmic ray density in the galaxy,
provided that up to 10\% of the available SN kinetic energy is used to
accelerate particles.
The energetic cosmic rays observed on earth are primarily nuclear cosmic
rays, unfortunately observations of SNRs lead primarily to the detection
of electron cosmic rays due to synchrotron or inverse Compton radiation.
In principle, nuclear cosmic rays can be detected, as collisions
of energetic heavy particles with background matter leads to the production
of pions. The neutral pions decay emitting two photons, which can be detected
by gamma-ray satellites such as the future GLAST mission or by TeV \u Cerenkov
telescopes.
Several SNRs, such as SN 1006\cite{tanimori98}, \casa\ \cite{aharonian01}, 
and \rxjSNR\ \cite{enomoto02},
have been detected with Cherenkov telescopes, 
but it is not clear whether  the observed
emission is due to inverse Compton emission from electrons or pion decay
due to nuclear cosmic rays.

The recent discovery of X-ray synchrotron emission from a number of
SNRs, SN 1006\footnote{
See the green-blueish color in the \xmm\ image of SN,1006,
Fig.~\ref{color} bottom left.} 
\cite{koyama95,allen01,dyer01}, 
\casa\ \cite{vink03a},
RCW~86 \cite{rho02}, 
\rxjSNR \cite{koyama97,slane99}, and G266.2-1.2 (``Vela jr'') 
\cite{slane01a},
have extended the observed electron energies from the GeV range (from
radio observations) to 10-100~TeV energies.
Although this is additional proof that electrons are accelerated to 
high energies, the observed energy range falls short of the ``knee'' 
energy \cite{reynolds99}.
Interestingly this list includes all remnants detected at TeV energies.

The identification of the TeV emission as either pion decay or inverse
Compton radiation depends on the magnetic field strength, which is needed
to convert the observed X-ray photon energies into electron energies.
This dependency can be used to infer the magnetic field strength assuming 
that the TeV emission is inverse Compton radiation \cite{dyer01}, or,
if a field strength is estimated, to identify the nature of the TeV emission
\cite{berezhko02}.

One usually assumes that the post shock magnetic field strength is
compressed by the shock and is at most a factor 4 times
the canonical galactic field value of $B\sim5~\mu$G. 
However, the spatial resolution of \chandra\ has lead to resolving the
narrow X-ray synchrotron structures at the shock fronts of \casa\ and
SN\,1006 \cite{vink03a,bamba03}, from which one can infer
the interior magnetic fields close to the shock front, if one
has an estimate of the shock velocity \cite{vink98a,delaney03},
as illustrated in Fig.~\ref{casarims}.
For both remnants the post shock magnetic field turns
out to be around 100~$\mu$G \cite{vink03a,berezhko03c}.
If correct this would mean that the TeV gamma-ray emission from SN 1006
is due to pion decay \cite{berezhko02}, 
whereas for \casa\ pion decay is likely, 
but inverse Compton emission cannot be excluded \cite{vink03a}.

The large magnetic fields found in \casa\ and SN\,1006 provide additional
support for the idea that SNRs can accelerate nucleons up to the ``knee''
energy, as the X-ray observations indicate that the electron maximum energy
is bound by synchrotron losses, not by acceleration time. If the latter
were the case then the maximum nucleon energy would be comparable
to the maximum electron energy, i.e. $< 100$~TeV. A higher magnetic field
allows for the possibility that nucleons are accelerated to higher energies,
and, moreover, higher magnetic fields increase the acceleration efficiency.

\begin{figure*}
\centerline{
  \psfig{figure=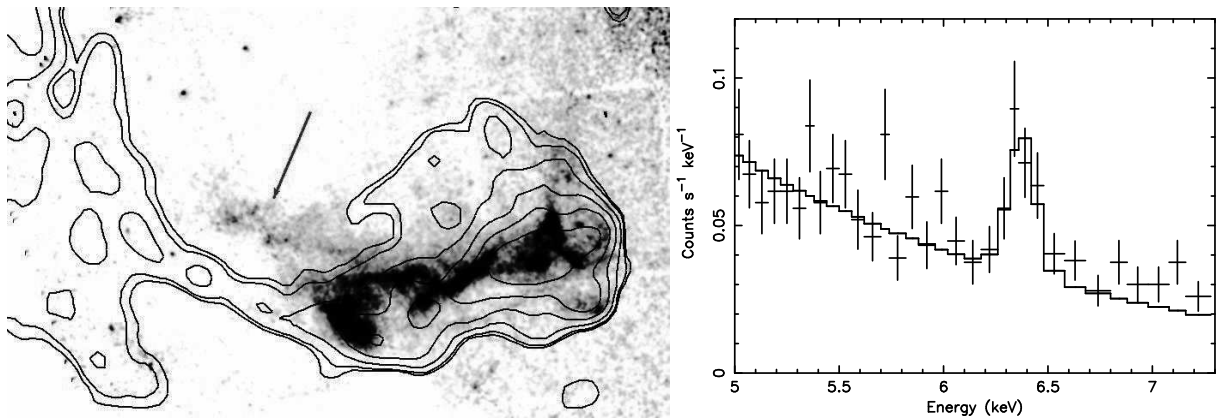,width=0.55\textwidth}
  \psfig{figure=jvink1_f7b.ps,angle=-90,width=0.45\textwidth}}
\caption{ \xmm\ image in the X-ray 2-7 keV band, which is
dominated by continuum emission (left), and has totally different
morphology than in softer X-ray bands. Radio contours are overlayed.
The arrow indicates a region of strong X-ray continuum without obvious
radio synchrotron emission.
The region that emits hard X-ray continuum also emits Fe-K emission
at 6.4~keV (right) \cite{vink97,vink02b,rho02}.
\label{rcw86}}
\end{figure*}

The detection of X-ray synchrotron emission from shell-type SNR has advanced
our knowledge of cosmic ray acceleration. However, one should approach
its detection with some caution. So far the identification of X-ray synchrotron
emission has been based on a power law continuum shape, and the (near)
absence of (thermal) line emission.
For \casa\ the detection of X-ray synchrotron emission was based on
the presence of hard X-ray continuum \cite{allen97}.
Its nature is, however uncertain. \xmm\ images of emission around 10~keV 
show that emission is still coming from the interior of \casa, whereas
one does not expect X-ray synchrotron emission from the inside,
as synchrotron losses due to 
the high interior magnetic field of $>0.5$~mG \cite{vink03a} make
it unlikely that a very energetic electron population can be maintained.
An alternative mechanism for the X-ray continuum in terms of non-thermal 
bremsstrahlung from electrons accelerated by lower hybrid waves has been
proposed \cite{laming01b} and model spectra fit the
\sax\ data well \cite{vink03a} (Fig.~\ref{ti44}).
Also for the hard X-ray emission from IC~443 has a non-thermal bremsstrahlung
origin been proposed \cite{bocchino00,bocchino03}. This emission
may arise from an interaction of the SNR with molecular clouds or from
fast moving ejecta fragments \cite{bykov02}.

Another case where there has been a controversy about the nature of the
X-ray continuum emission concerns the X-ray continuum from RCW 86,
a 40\arcmin\ sized SNR (Fig.~\ref{color}).
\asca\ and \sax\ data showed that 
the soft X-ray emission, dominated by line emission, has a very different
morphology than the continuum emission, 
which manifests itself most clearly above 2~keV \cite{vink97,bocchino00}
(Fig.~\ref{color}). 
This was taken as evidence for X-ray synchrotron
emission \cite{borkowski01}, in analogy with SN\,1006. 
However, the hard X-ray continuum is not entirely featureless,
since Fe-K line emission has been solidly detected and seems to come from
the hard X-ray emitting regions.
So an alternative theory for the hard X-ray emission is that it is due
to (non-thermal) bremsstrahlung from electrons with energies up to 100~keV,
which are also responsible for the Fe-K emission \cite{vink97,vink02b}.
Additional evidence for questioning the synchrotron hypothesis
is the lack of radio synchrotron emission from hard X-ray producing regions
(Fig.~\ref{rcw86}).
The Fe-K emission has an energy of 6.4~keV, indicating that is caused
by low ionization stages of Fe, i.e. either from a cold or a severely 
underionized plasma.
The problem with the bremsstrahlung hypothesis is that the presence of
the bremsstrahlung producing electrons has to heat up the background material,
which should manifest itself in line emission of O VII \cite{rho02}.
The synchrotron hypothesis, on the other hand, needs the presence of an 
additional hot, but very underionized pure Fe plasma, in order to 
explain the Fe-K emission \cite{rho02}.
This debate will be helped by deeper \chandra\ and \xmm\ observations.
Especially observing the spatial morphology in Fe-K emission is important,
because 
the bremsstrahlung hypothesis predicts a morphology similar to the continuum
emission, and the synchrotron plus hot pure Fe hypothesis predicts patches
of Fe-K emission and spatially uncorrelated, diffuse X-ray continuum emission.

Whatever the outcome it will either give new
insights in the acceleration of electrons at the low end, or at the high end
of the electron cosmic ray spectrum. Moreover, if the Fe-K is indeed from
pure Fe this may have interesting consequences for our understanding
of RCW 86's progenitor.

\section{Concluding remarks}
The last paragraph shows that to some extent the issue of  cosmic
ray acceleration and explosive nucleosynthesis are intertwined.
In order to assess abundance patterns an understanding of the
radiation processes, bremsstrahlung, line emission, and synchrotron emission
is needed. The possible presence of synchrotron emission in a line
dominated spectrum has consequences for the mass estimates \cite{favata97}.
The discovery of X-ray synchrotron radiation from shell-type remnants
have complicated the analysis of their X-ray emission, but has given
unexpected new insights in the particle acceleration properties of
supernova remnants.

These advances in our understanding have been made possible by 
the emergence of spatially resolved X-ray spectroscopy, first
with \asca\ and \sax, and now fruitfully continued with \chandra\ and \xmm.

\vskip 0.1cm
I thank Kurt van der Heyden for generating Fig.~\ref{typeIa}.
This work was supported by NASA's
Chandra Postdoctoral Fellowship Award No. PF0-10011
issued by the Chandra X-ray Observatory Center, which is operated by the
SAO under NASA contract NAS8-39073.



\end{document}